\pdfoutput=1
\documentclass[a4paper]{article}
\usepackage{spconf,amsmath,graphicx,booktabs,amssymb,subfigure,threeparttable,multirow,float,hyperref,url,color,threeparttable,bbding,textpos}
\hypersetup{
    colorlinks=true,
    linkcolor=red,
    filecolor=red,      
    urlcolor=red,
    citecolor=green,
}

\DeclareMathOperator {\round}{round}
\DeclareMathOperator {\sisdr}{SI-SDR}

\title{LIMUSE: LIGHTWEIGHT MULTI-MODAL SPEAKER EXTRACTION}
\name{Qinghua Liu$^1$$^,$$^2$$^\dag$\thanks{$^\dag$Co-first authors}, Yating Huang$^1$$^,$$^3$$^\dag$, Yunzhe Hao$^1$$^,$$^3$, Jiaming Xu$^1$$^,$$^3$$^\ddag$\thanks{$^\ddag$Corresponding authors}, Bo Xu$^1$$^,$$^3$$^,$$^\ddag$}
\address{
$^1$Institute of Automation, Chinese Academy of Sciences (CASIA), Beijing, China\\
$^2$Qiushi Honors College, Tianjin University, Tianjin, China\\
$^3$School of Future Technology, University of Chinese Academy of Sciences, Beijing, China}

\begin{document}
\ninept
\maketitle
\begin{abstract}
Multi-modal cues, including spatial information, facial expression and voiceprint, are introduced to the speech separation and speaker extraction tasks to serve as complementary information to achieve better performance. However, the introduction of these cues brings about an increasing number of parameters and model complexity, which makes it harder to deploy these models on resource-constrained devices. In this paper, we alleviate the aforementioned problem by proposing a Lightweight Multi-modal framework for Speaker Extraction~(LiMuSE).  We propose to use GC-equipped TCN, which incorporates Group Communication~(GC) and Temporal Convolutional Network~(TCN) in the Context Codec module, the audio block and the fusion block. The experiments on the MC\_GRID dataset demonstrate that LiMuSE achieves on par or better performance with a much smaller number of parameters and less model complexity. We further investigate the impacts of the quantization of LiMuSE.
Our code and dataset are provided$\footnote{https://github.com/aispeech-lab/LiMuSE}$.
\end{abstract}
\noindent\textbf{Index Terms}: Model Compression, Speaker Extraction, Multi-Modality, Group Communication, Quantization

\begin{textblock*}{\textwidth}(0cm,11cm)
\tiny
\noindent
Copyright 2022 IEEE. Published in the 2022 IEEE Spoken Language Technology Workshop (SLT) (SLT 2022), scheduled for 19-22 January 2023 in Doha, Qatar. Personal use of this material is permitted. However, permission to reprint/republish this material for advertising or promotional purposes or for creating new collective works for resale or redistribution to servers or lists, or to reuse any copyrighted component of this work in other works, must be obtained from the IEEE. Contact: Manager, Copyrights and Permissions / IEEE Service Center / 445 Hoes Lane / P.O. Box 1331 / Piscataway, NJ 08855-1331, USA. Telephone: + Intl. 908-562-3966.
\end{textblock*}

\section{Introduction}
\label{sec:intro}
The Cocktail Party Problem is a fundamental topic in auditory scene analysis, which refers to the phenomenon that humans are able to focus his/her attention on specific auditory stimuli by masking out the acoustic background noise in complex auditory scenes~\cite{bronkhorst2000cocktail}.
With the rapid development of deep learning technologies, there is an emerging trend of using deep-learning-based methods to model the Cocktail Party Problem.
In the speech processing community, these approaches can be categorized into two types: blind speech separation and target speaker extraction. Blind speech separation aims to mimic the human’s bottom-up stimulus-driven attention by simultaneously estimating the speech of different speakers from the mixture, which is often confronted with two problems: permutation problem~\cite{yu2017permutation} and output dimension mismatch problem~\cite{Chen2017DeepAN}.
By contrast, speaker extraction (also termed as target speech separation) only extracts the speech of interest from the overlapped audio of simultaneous speakers. It emulates humans’ top-down voluntary focus~\cite{xu2020spex} by utilizing the information from the target speaker as prior knowledge to direct the attention to the interested speaker. In this way, speaker extraction avoids the permutation problem and output mismatch problem inherently.

Speaker extraction often requires additional information that identifies the target speaker. 
Voiceprint, which characterizes a speaker, has proved to be an important cue in auditory scene analysis~\cite{Wang2020,xu2020spex,ge2021multi,hao2021wase}. 
Recently, incorporating visual information such as face or lip image sequences, in which phonetic or prosaic content coincides with muscle movements, into speaker extraction systems becomes an emerging research focus to improve the robustness and separation quality~\cite{wu2019time,lu2019audio,ephrat2018looking,zhang2021avss}.
What's more, multi-channel-based methods can benefit from additional direction information to further improve the performance~\cite{li2021speaker,li2019direction}. 
By utilizing all the available information of the target speaker mentioned above, 
a multi-channel multi-modal system is proposed~\cite{gu2020multi} and achieves superb performance.
In challenging acoustic environments, the acoustic information of the target speaker can be blurry. The information of other modalities serves as complementary and steady information to increase the quality of the interested speech. 
However, the addition of multi-modal information, particularly visual information, typically results in larger model sizes and higher computational costs, which poses severe challenges for the deployment of these deep-learning-based models on resource-constrained devices, such as mobile phones and other edge devices.

In order to alleviate the deployment problems on resource-constrained devices, there are mainly two types of methods: design of novel model architectures and model compression methods.
MobileNet~\cite{howard2017mobilenets} proposes to use depth-wise separable convolution to decrease the parameters of the traditional convolutional layer. MobileNetV2~\cite{sandler2018mobilenetv2} further proposes linear bottleneck and inverted residual for lightweight model structure design.
The practice of Group Communication~\cite{luo2021ultra} proves that a small, group-shared module is adequate to preserve the model capacity given the inter-group modeling step.
Furthermore, Group Communication with Context Codec (GC3)~\cite{luo2021group} cultivates a more efficient model design to decrease model size without sacrificing performance.
Besides, various model compression methods have been proposed to lighten deep neural networks, such as parameter pruning~\cite{luo2017thinet,han2015learning}, knowledge distillation~\cite{hinton2015distilling,heo2019knowledge} and quantization~\cite{rastegari2016xnor,wu2018training,zhou2016dorefa}. 
In contrast to the adjustments of model architectures, the quantization technology compresses a model by converting a full-precision neural network into a low-bitwidth integer version.
By limiting the model topology and quantizing the model to 8-bit integer form, VoiceFilter-Lite~\cite{Wang2020} makes the model tiny and fast to meet strict on-device production requirements. TinyWASE~\cite{tinywase2022} investigates the effects of quantization function in audio-only speaker extraction tasks and greatly compresses the model size without sacrificing much performance.

\begin{figure*}[htb]
	\centering
	\includegraphics[width=\linewidth]{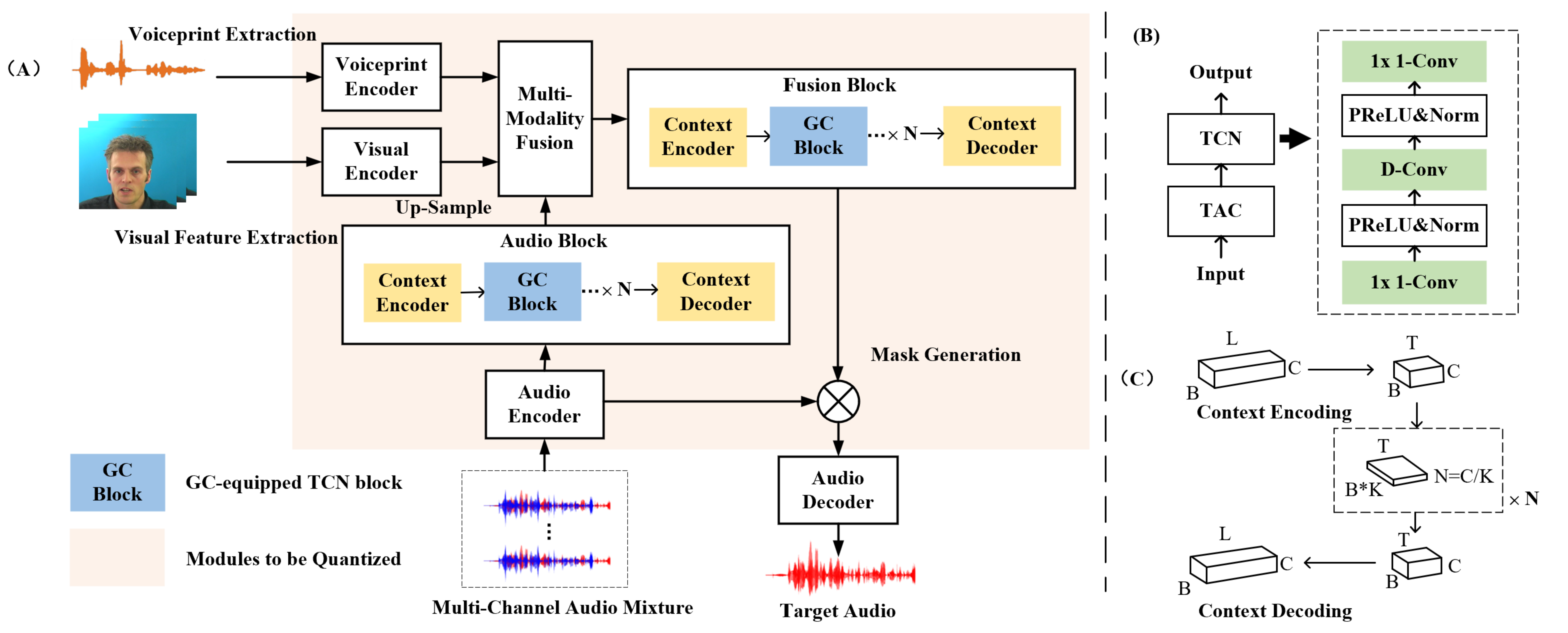}
	\caption{An illustration of LiMuSE framework. (A) The overall architecture of the proposed model and modules to be quantized. `GC Block' indicates the GC-equipped TCN block. (B) Details of GC-equipped TCN block. `D-Conv' indicates a depth-wise convolution operation. (C) Sequence processing pipeline in the Audio Block and Fusion Block. `B' denotes batch size, `C' denotes number of feature channels, `L' denotes original length of temporal sequence and `T' denotes compressed temporal sequences.}
	\label{fig:flowchart}
\end{figure*}

In this paper, we propose a lightweight multi-modal speaker extraction framework, which incorporates multi-channel information, target speaker's visual feature and voiceprint as reference information and further apply Group Communication, Context Codec and an ultra-low bit quantization technology to reduce the model size and complexity while maintaining relatively high performance.
Our contributions can be concluded as follows:

\begin{enumerate}
\item As far as we know, it is the first time that Group Communication has been explored in a multi-modal speaker extraction model. The results have shown the effectiveness, and an appropriate number of groups may promote the inter-modal information exchange and even achieve improvements;
\item We incorporate GC-equipped TCN blocks in the process of sequence modeling and Context Codec to decrease both the width of neural network and length of sequential feature without using RNN-related structures, which is more friendly to incorporate quantization technologies, as attempts for quantization of RNNs may show considerable performance degradation with low bit-width weights and activations;
\item We further compress the model size with quantization technologies. The results have shown that our model with ultra-low bit quantization can also maintain a comparable performance, compared with the baselines.
\end{enumerate}

\section{Model Design}
\label{sec:design}
Our proposed model is a multi-stream architecture that takes multi-channel audio mixture, target speaker's enrolled utterance and visual sequences of detected faces as inputs, and outputs the target speaker's mask in the time domain. The encoded audio representations of the mixture are then multiplied by the generated mask to obtain the target audio. Figure~\ref{fig:flowchart}
provides an overview of our network.

\subsection{Audio and Video Representations}
\subsubsection{Audio Stream}
\label{sssec:audio}
The audio encoder encodes the multi-channel audio mixture into spectrum-like representation in the time domain, and the audio decoder is used to transform the masked encoder representation to the target waveform. The encoder and the decoder follow the same structure in \cite{luo2019conv}, which uses a 1-dimensional convolutional layer and a 1-dimensional transpose convolutional layer, respectively.

Formulating the audio signals in the time domain has some advantages over the traditional time-frequency representations obtained by Short-Time Fourier Transform (STFT), including trainable weights, finer coding granularity and avoidance of phase reconstruction~\cite{hao2021wase}.

\subsubsection{Speaker Embedding Stream}
\label{sssec:voiceprint}
The voiceprint models the characteristics of the target speaker and serves as prior knowledge to help the model filter out the target speaker's speech from the mixture.
The voiceprint encoder takes the speaker embeddings of the target speaker as input. In this paper, we use an off-the-shelf speaker diarization toolkit pyannote~\cite{Bredin2020} to extract the speaker embeddings $\mathbf{s_0}\in\mathbb{R}^{U}$, where $U$ denotes the length of the speaker embedding feature dimension. 

In order to align with the time dimension of the mixed audio stream, we stack the speaker embeddings along the time dimension to form $\tilde{S}=[\dots,s_t,\dots]\in\mathbb{R}^{U \times T}$. 
The voiceprint encoder uses a fully connected (FC) layer to map the feature dimension of the voiceprint to obtain $\mathbf{S}\in\mathbb{R}^{N \times T}$, which is the same shape as the representation of the audio mixture stream.

\subsubsection{Video Stream}
\label{sssec:video}
We obtain face embeddings of the target speaker from videos using the network in \cite{zhou2018talking}, which extracts speech-related visual features from visual inputs explicitly by the adversarially disentangled method. The face embedding extraction network first learns joint visual and auditory representation from clean audio and video pairs to establish the correlation between the audio features and the visual features, and then disentangles the speech-related visual features from the joint audio-visual representation with the adversarial training method.

The pretrained face embedding extraction network is trained on LRW dataset~\cite{Joon2017lip} and MS-Celeb-1M dataset~\cite{MSCeleb1M2016Guo} to obtain $\mathbf{\tilde{V}}\in\mathbb{R}^{D \times E}$, where $D$ denotes the feature dimension of the face embedding and $E$ denotes the number of frames in each video stream. The visual encoder then maps the extracted face embeddings to match the feature dimension of the encoded audio features and obtain $\mathbf{v_0}\in\mathbb{R}^{N \times E}$. Moreover, since the time resolution of the video stream and the audio stream is different, we upsample the face embeddings along the temporal dimension to synchronize the audio stream and the video stream by nearest neighbor interpolation to obtain $\mathbf{V}\in\mathbb{R}^{N \times T}$. The face embeddings are then used to model the temporal synchronization and interaction between visemes and speech~\cite{bear2017phoneme}.

In this paper, we simply use a fully connected layer to map the face embedding rather than depth-wise separable convolution adopted by AV-ConvTasnet~\cite{wu2019time} or LSTM~\cite{yu2019review} adopted by AVMS~\cite{zhang2021avss}.

\subsection{Separation Module}
\subsubsection{GC-Equipped TCN block}
\label{ssec:gc}
In order to reduce the number of parameters, we apply Group Communication (GC)~\cite{luo2021group} to the separation modules. Given a feature $\mathbf{H}\in\mathbb{R}^C$, it can be decomposed into $K$ groups of low-dimensional feature vectors $\{\mathbf{g}^i\}_{i=1}^K$. A GC module is applied to capture the inter-group dependencies.

We use the Transform-Average-Concatenate (TAC) module~\cite{luo2020end} for the GC module. It is similar to the squeeze-and-excitation design paradigm~\cite{hu2018squeeze}, where there is a residual connection between a pool of features and a global-averaged feature which is generated to summarize the input feature pool. 
For ${\{ \mathbf{g}^i \} }_{i=1}^K$, a fully-connected layer followed by PReLU activation~\cite{he2015delving} is used for the transformation step to generate $\mathbf{f}^i$. Then all $\mathbf{f}^i$s are averaged and passed through another fully connected layer with PReLU for the average step to get $\hat{\mathbf{f}}^i$. Finally, $\hat{\mathbf{f}}^i$ is concatenated with $\mathbf{f}^i$ and passed to the third fully connected layer with PReLU to generate $\hat{\mathbf{g}}^i$. A residual connection is added between $\mathbf{g}^i$ and $\hat{\mathbf{g}}^i$. 
It is demonstrated in~\cite{luo2021group} that using TAC for the Group Communication module achieves better performance and lower computational costs than other network structures, such as BLSTM and self-attention.

In this paper, Group Communication is applied before each Temporal Convolutional Network (TCN) block~\cite{luo2019conv}, which forms a GC-equipped TCN block, as is shown in Figure~\ref{fig:flowchart} (B). 
Each TCN block consists of a depth-wise separable convolution, PReLU activation~\cite{he2015delving} and a layer normalization operation.
By introducing Group Communication to the TCN block, we can reduce the model width for the TCN block by $K$ (number of groups) times when there is no overlapping within the groups.

\subsubsection{Context Codec with GC-Equipped TCN block}
On one hand, 
speech has a rich temporal structure over multiple time scales~\cite{toledano2018multi}.
On the other hand, 
shorter kernel length of the convolution in the audio encoder leads to better separation performance and higher model complexity at the same time. Motivated by the above two characteristics, Context Codec~\cite{luo2021group} is proposed to decrease sequence length, which is similar to a nonlinear downsampling step by compressing the context of feature vectors into a summary one and then decompress it back to the context.
A Context Codec module typically contains a context encoder and a context decoder. In this paper, instead of using RNN~\cite{luo2021group}, we propose to use the GC-equipped TCN network for the Context Codec module.

To be specific, given a sequence of feature $\mathbf{H}\in\mathbb{R}^{C \times L}$, where $C$ denotes the number of feature channels, the context encoder first splits it along the temporal dimension into blocks ${\{ \mathbf{D}^i \} }_{i=1}^T \in\mathbb{R}^{C \times S}$ with an overlap ratio of 50\%, where $T$ denotes the number of context blocks and $S$ denotes the context size as illustrated in Figure~\ref{fig:flowchart}~(C).
Then the context encoder applies a GC-equipped TCN network (Section~\ref{ssec:gc}) on each $\mathbf{D}^i$ to establish inter-group correlation and obtains ${\hat{\mathbf{D}}}^i$. At last, a mean operation is applied on the context dimension of $\hat{\mathbf{D}}$ to produce a sequence of summarization vectors ${\{ \mathbf{p}^i \} }_{i=1}^T$ which represent the whole feature sequence. 
The compressed sequence $P \triangleq {\{ \mathbf{p}^i \} }_{i=1}^T \in\mathbb{R}^{C \times T}$ can then be sent into feature processing module with $T \ll L$. 
The context decoder adds the output of feature processing module to each time step in $\mathbf{D}^i$ and applies another GC-equipped TCN network for the nonlinear transformation. Overlap-and-add is then applied on ${\{ \hat{\mathbf{D}}^i\}}_{i=1}^T$ to form the sequence of feature of the original length ${ \hat{\mathbf{H}}} \in\mathbb{R}^{C \times L}$. The implementation details can be found in our released source code.

\subsubsection{Audio Block and Fusion Block}
In order to capture the long-term dependencies in the encoded audio features, the audio block is composed of several GC-equipped TCN blocks. In the audio block, we use two repeats of the TCN networks, and each repeat of the TCN network is composed of several TCN blocks. The dilation factor of depth-wise convolution increases exponentially by two in each repeat, which enlarges the receptive field of the module to capture the correlation in a larger time scale.

The output of the audio block is gathered with the encoded voiceprint and visual features through concatenation. Then we use a repeat of GC-equipped TCN blocks with an exponential increasing dilation factor of depth-wise convolution in the fusion block to process the fused features. The voiceprint and visual cues serve as auxiliary information to help the model extract the target speech from the mixture. A convolutional layer is followed by the fusion block to generate the mask of the target speaker.

\subsection{Ultra-low bit Quantization}
\label{sec:quantization}
In this section, we further investigate how to apply quantization strategies~\cite{tinywase2022} for multi-modal scenarios.
\subsubsection{Weight Quantization}
\label{ssec:weight}
For weight quantization, we propose to use the quantization function~\cite{yang2019quantization} to quantize weights to ultra-low bits for multi-modal speaker extraction tasks. The quantization function combines a linear combination of several simple functions to reformulate the ideal quantization operation and gradually approximates it by gradually increasing the temperature parameter during training. 

Suppose that $x$ is the full-precision value to be quantized, $y$ is the quantized integer constrained to a predefined set $\gamma=\{\gamma_1, \gamma_2, \dots, \gamma_n\}$. For example, if we perform 3-bit weight quantization, then $y$ can be defined as $\gamma=\{-3, -2, -1, 0, 1, 2, 3\}$.
The ideal quantization operation can be reformulated as a combination of unit step functions as follows:
\begin{equation}
\label{eq:ideal}
y = \sum_{i=1}^n{s_i A (\beta x - b_i)-o},
\end{equation}

\begin{equation}
\label{eq:unit}
A(x)=\left \{
\begin{array}{rcl}
1      & {x \geq 0,}\\
0      & {x < 0,}
\end{array} 
\right.
\end{equation}
where $n=|\gamma|-1$, $\beta$ is the scale factor of the input. Eq~\ref{eq:unit} defines the unit step function, where  $s_i={\gamma}_{i+1}-{\gamma}_{i}$ and $b_i$ is the bias for the unit step function. The global offset is defined as $o=\frac{1}{2}\sum_i^ns_i$.

However, the unit step function in Eq~\ref{eq:unit} is non-differentiable and thus can not be trained from end to end. The quantization function replaces the unit step function with differentiable sigmoid function in the training procedure as follows: 
\begin{equation}
y = Q(x) = \alpha \left( \sum_{i=1}^n{s_i\sigma (T(\beta x - b_i))-o} \right),
\end{equation}
\begin{equation}
\sigma (T x) = \frac{1}{1+\exp(-Tx)},
\end{equation}
where $T$ denotes the temperature parameter, $\alpha$ and $\beta$ are learnable scale factors of the output and input, respectively. While in the inference phase, the ideal quantization function in Eq~\ref{eq:ideal} is used. Performance may suffer if different quantization functions are used throughout the training and inference phases. To narrow the gap, temperature $T$ is introduced to the sigmoid function. When $T$ is large, the gap between the sigmoid function and the unit step function is minimal, and vice versa. Starting from a small temperature and gradually increasing it, the quantized neural networks can be well learned. 

In this paper, we perform layer-wise non-uniform quantization, in which weights from different layers use different quantization functions. We quantize all neural network modules except the audio decoder. We also do not quantize the PReLU activation and the layer normalization function because they may be sensitive to quantization and have a small number of parameters.
To perform non-uniform quantization, we use K-means clustering on full-precision weights to obtain $n+1$ ranked cluster centers $c=\{c_1, c_2, \dots, c_{n+1}\}$ in the ascending order. The bias $b_i$ is initialized as $b_i=\frac{c_i+c_{i+1}}{2}$ and gets fixed during training.

\subsubsection{Activation Quantization}
\label{ssec:activation}
We also quantize the activations of the corresponding quantized layers to make matrix multiplication run faster. For activation quantization, we use the most commonly used min-max linear quantization~\cite{krishnamoorthi2018quantizing}, which is defined as:
\begin{equation}
Q(x) = \round \left( \frac{x-\mathrm{X}_{\mbox{\tiny{min}}}}{s} \right),
\end{equation}
\begin{equation}
s = \frac{\mathrm{X}_{\mbox{\tiny{max}}}-\mathrm{X}_{\mbox{\tiny{min}}}}{2^p-1},
\label{eq:activation}
\end{equation}
where $\mathrm{X}$ is the tensor to be quantized, $x$ is the element in $\mathrm{X}$, $\mathrm{X_{min}}$ and $\mathrm{X_{max}}$ are the minimum and maximum elements, respectively. In this paper, the activation quantization bits are $8$, then $p = 8$ in equation~\ref{eq:activation}.
Straight-Through Estimator (STE)~\cite{courbariaux2015binaryconnect} is used to backpropagate the gradients through quantized activations.

\subsection{Loss Function}
Scale-Invariant Signal-to-Distortion ratio (SI-SDR) between the predicted wave $\hat{\mathbf{s}}$ and the clean wave $\mathbf{s}$ is used to optimize the network from end to end. The loss function is defined as:
\begin{align}
  \label{eq:SI-SDR}
  \begin{cases}
  \mathbf{s}_{target}=\frac{\left \langle {\hat{\mathbf{s}},\mathbf{s}} \right \rangle \mathbf{s}}{{\left \| \mathbf{s} \right \|}^2};\\
  \mathbf{e}_{noise}=\hat{\mathbf{s}}-\mathbf{s}_{target};\\
  \sisdr(\mathbf{s}_{target}, \mathbf{s})=10 \log_{10} {\frac{{\left \| \mathbf{s}_{target} \right \|}^2}{{\left \| \mathbf{e}_{noise} \right \|}^2}}; \\
  \end{cases}
\end{align}

\section{EXPERIMENT SETUP}
\label{sec:exp}

\subsection{Datasets}
We evaluate our system on speaker extraction problems using the GRID dataset~\cite{cooke2006audio}, which contains 18 male and 15 female speakers, and each of them has 1,000 frontal 3-second face video recordings. The sampling rate of speech is 16 kHz. And the videos are sampled at 25 FPS. We randomly select 3 males and 3 females to construct a validation set of 2.5 hours and another 3 males and 3 females for a test set of 2.5 hours. The rest of the speakers form the training set of 30 hours. 

To construct a 2-speaker mixture, we randomly choose two different speakers and select audio from each chosen speaker, then mix these two clips at a random SNR level between -5 dB and 5 dB. Then we use the SMS-WSJ toolkit~\cite{SmsWsj19} to generate the simulated anechoic dual-channel audio mixture. Specifically, we place 2 microphones at the center of the room, and the distance between the microphones is 7 cm. 
The implementation code along with the Multi-Channel GRID (MC\_GRID) dataset and the feature embeddings obtained from raw videos and reference speech used in our experiments are all available.

\subsection{Training Details}
We first train the full-precision network for 50 epochs with the initial learning rate set to $1e^{-3}$. Then, to train the quantization network, we load the weights of the pretrained full-precision model to get a good starting point. In the quantization stage, the whole separation module and mask generation module are quantized. The weights of the encoders are also quantized. 
In particular, the weights and the activations are quantized to 3 bits and 8 bits, respectively. 
The temperature $T$ in the quantization function is set to 5 in the beginning and increases linearly with respect to the epochs, which is $T=5 \times epochs$. The learning rate is halved if the accuracy on the validation set does not improve for 4 consecutive epochs. 
The training process would stop if the accuracy on the validation set does not improve for 6 consecutive epochs. 
Adam~\cite{kingma2015adam} is used as the optimizer. 
Gradient clipping with maximum L2-norm of 5 is applied during training. 

The hyperparameters of the proposed neural network are shown in Table~\ref{tb:symdes}.

\begin{table}[htbp]
\centering
\caption{Hyperparameters of LiMuSE.}
\begin{tabular}{ccc}
\hline
\textbf{Symbol} & \textbf{Description} & \textbf{Value} \\ \hline
$N$ & Number of channels in audio encoder & 128 \\
$L$ & Length of the filters (in audio samples) & 32\\
$P$ & Kernel size in convolutional blocks & 3 \\
$R_a$ & Number of repeats in audio block & 2 \\
$R_f$ & Number of repeats in fusion block & 1 \\
$S$ & Context size (in frames) & 32\\
$W_q$ & Weight quantization bits & 3 \\
$A_q$ & Activation quantization bits & 8 \\
$T_0$ & Temperature increment per epoch & 5  \\ \hline
\end{tabular}
\label{tb:symdes}
\end{table}

\begin{table*}[htbp]
\centering
\caption{Comparison between LiMuSE and other speaker extraction (SE) and speech separation (SS) baselines on GRID. Input streams include audio (A), voiceprint (Vp) and visual feature (Vis).}
\begin{threeparttable}
\begin{tabular}{ccccccc}
\hline
\textbf{Method} & \textbf{Task} &\textbf{Input Stream} & \textbf{SI-SDRi (dB)} & \textbf{SDRi~(dB)} & \textbf{Model Size}  & \textbf{MACs} \\ \hline
LiMuSE  & SE & A+Vp+Vis & 17.25 & 17.50 & 0.48MB & 7.52G  \\
MuSE~\cite{pan2021muse}\tnote{*} & SE & A+Vis & 8.53 & - & 54.27MB & 25.87G \\
AVMS~\cite{zhang2021avss} & SS & A+Vis & - & 15.74 & 22.34MB & 60.66G \\ 
AVDC~\cite{lu2019audio} & SS & A+Vis & - & 8.88 & - & - \\ 
Conv-TasNet~\cite{luo2019conv}\tnote{**} & SS & A & 14.98 & 15.27 & 13.28MB & 21.44G \\ \hline
\end{tabular}
\label{tb:result}

\begin{tablenotes}
\footnotesize
\item[*] This paper only reports cross-dataset evaluation results on the GRID dataset. It is trained on VoxCeleb2~\cite{chung2018voxceleb2} and tested on GRID.
\item[**] We re-implement and train Conv-Tasnet on GRID for comparison.
\end{tablenotes}
\end{threeparttable}
\end{table*}

\section{RESULTS AND DISCUSSIONS}
\label{sec:result}
\subsection{LiMuSE vs baselines on GRID}

\label{baseline}
We report SI-SDR improvement (SI-SDRi) and SDR improvement (SDRi) scores for the evaluation of the model performance. For model complexity, the model size and the number of MAC operations (MACs) are reported as metrics. MACs are calculated using an open-source
toolbox\footnote{https://github.com/Lyken17/pytorch-OpCounter}.
As is shown in Table~\ref{tb:result}, LiMuSE exhibits competitive performance compared with other existing multi-modal speech separation and speaker extraction baselines with substantially smaller model size and model complexity. 

\begin{table}[htbp]
\centering
\caption{Comparison between different visual front-ends. The inputs to the visual front-ends are the face sequences of the speakers. Performance is evaluated on LiMuSE without quantization. K stands for the number of groups.}
\begin{tabular}{ccccc}
\hline
\textbf{Method} &  \textbf{K} & \textbf{SI-SDR (dB)} & \textbf{SDR (dB)} \\ \hline
\multirow{2}{*}{Adversarially Disentangle} & 32 & 21.75 & 22.61  \\
 & 16 & 24.27 & 24.83 \\ \hline
\multirow{2}{*}{Deep AVSR~\cite{afouras2018deep}} & 32 & 18.87 & 20.19 \\
 & 16 & 20.98 & 22.02\\ \hline
\end{tabular}
\label{tb:front}
\end{table}

\begin{table}[htbp]
\centering
\caption{Comparison between different visual encoder modules. Performance is evaluated on LiMuSE without quantization. K stands for the number of groups.}
\begin{tabular}{cccccc}
\hline
\textbf{Method} &  \textbf{K} & \textbf{SI-SDR~(dB)} & \textbf{SDR~(dB)} & \textbf{\#Params} \\ \hline
\multirow{2}{*}{FC layer} & 32 & 21.75 & 22.61 & 0.41M  \\
 & 16 & 24.27 & 24.83 & 1.12M \\ \hline
\multirow{2}{*}{V-TCN} & 32 & 22.81 & 23.20 & 0.74M \\
 & 16 & 25.12 & 25.84 & 1.46M\\ \hline
\end{tabular}
\label{tb:visualenc}
\end{table}

\begin{table}[htbp]
\centering
\caption{Comparison between causal and non-causal settings. Performance is evaluated on LiMuSE without quantization.}
\begin{tabular}{cccccc}
\hline
\textbf{K} & \textbf{Causal} & \textbf{SI-SDR~(dB)} & \textbf{SDR~(dB)} & \textbf{SDRi~(dB)} \\ \hline
\multirow{2}{*}{32} & $\times$  & 21.75 & 22.61 & 22.40  \\
 & \checkmark & 20.59 & 21.64 & 21.43 \\ \hline
\multirow{2}{*}{16} & $\times$  & 24.27 & 24.83 & 24.63 \\
 & \checkmark & 22.23 & 22.93 & 22.72\\ \hline
\end{tabular}
\label{tb:causal}
\end{table}

\begin{table*}[htbp]
\centering
\caption{The effects of lightweight designs for LiMuSE on MC\_GRID. K stands for the number of groups. Q refers to quantization, GC refers to Group Communication, CC refers to Context Codec. Specifically, LiMuSE~(vanilla) refers to the model without using Group Communication, Context Codec and Quantization.}
\begin{tabular}{ccccccccc}
\hline
\textbf{Model} & \textbf{K} & \textbf{SI-SDR~(dB)} & \textbf{SDR~(dB)} & \textbf{SDRi~(dB)} & \textbf{\#Params} & \textbf{Model Size} & \textbf{MACs} \\ \hline
LiMuSE~(vanilla) & - & 23.54 & 24.02 & 23.83 & 8.95M & 34.14MB (100\%) & 53.34G (100\%)\\ \hline
\multirow{2}{*}{LiMuSE~(GC)} & 32 & 19.17 & 20.71 & 20.50 & \textbf{0.37M} & 1.40MB (4.1\%) & 5.94G (11.1\%) &  \\
 & 16 & 23.78 & 23.65 & 23.45 & 0.97M & 3.70MB (10.8\%) & 11.77G (22.1\%) \\ \hline
\multirow{2}{*}{LiMuSE~(GC+CC)} & 32 & 21.75 & 22.61 & 22.40 & 0.41M & 1.55MB (4.54\%) & \textbf{3.98G (7.46\%)} &  \\
 & 16 & \textbf{24.27} & \textbf{24.83} & \textbf{24.63} & 1.12M & 4.28MB (12.5\%) & 7.52G (14.1\%) \\ \hline
 \multirow{2}{*}{LiMuSE~(GC+CC+Q)} & 32 & 15.53 & 16.67 & 16.46 & 0.41M & \textbf{0.19MB (0.56\%)} & \textbf{3.98G (7.46\%)} &  \\
 & 16 & 17.25 & 17.71 & 17.50 & 1.12M & 0.48MB (1.41\%) & 7.52G (14.1\%) \\ \hline
\end{tabular}
\label{tb:ablation}
\end{table*}

\begin{table}[htbp]
\centering
\caption{Comparison between different weight quantization bits. The activation quantization bits are 8 in both settings.}
\begin{tabular}{cccc}
\hline
\textbf{$W_q$(bit)} &  \textbf{K} & \textbf{SDR (dB)} & \textbf{Model Size (MB)} \\ \hline
\multirow{2}{*}{3} & 32 & 16.67 & 0.19 \\
 & 16 & 17.71 & 0.48\\ \hline
\multirow{2}{*}{4} & 32 & 18.30 & 0.24  \\
 & 16 & 18.86 & 0.61 \\ \hline
\end{tabular}
\label{tb:compare_q}
\end{table}

\begin{table}[htbp]
\centering
\caption{The effects of the voiceprint cues and the visual cues for unquantized LiMuSE~(GC+CC) on MC\_GRID. K stands for the number of groups. -Vp refers to without using the voiceprint cue and -Vis refers to without using the visual cue.}
\begin{tabular}{ccccc}
\hline
\textbf{Model} & \textbf{K} & \textbf{SI-SDR~(dB)} & \textbf{SDR~(dB)} & \textbf{SDRi~(dB)} \\ \hline
 LiMuSE & 32 & 21.75 & 22.61 & 22.40 \\
 (GC+CC)& 16 & \textbf{24.27} & \textbf{24.83} & \textbf{24.63} \\ \hline
 LiMuSE & 32 & 19.17 & 20.71 & 20.50\\
 (GC+CC-Vp) & 16 & 21.13 & 22.38 & 22.17 \\ \hline
 LiMuSE & 32 & 14.75 & 16.32 & 16.11 \\
 (GC+CC-Vis) & 16 & 18.57 & 20.75 & 20.54 \\ \hline
\end{tabular}
\label{tb:ablation_cue}
\end{table}

\subsection{Comparison with different visual feature processing modules}
\label{compare_v}
We compare our visual feature processing module to those used in current audio-visual speech separation or speaker extraction models. For the pretrained visual front-end, we use the entire face region as model inputs. As is shown in Table~\ref{tb:front}, the Adversarily Disentangle approach outperforms Deep AVSR~\cite{afouras2018deep} (includes a 3D convolutional layer followed by a 18-layer ResNet~\cite{he2016deep}) in terms of face embedding performance.
For the visual encoder modules, we replace the widely adopted temporal convolutional block (V-TCN), which consists of 5 residual connected rectified linear unit (ReLU), batch normalization (BN) and depth-wise separable convolutional layers (DS Conv1D)~\cite{wu2019time,pan2021muse} with a simple fully connected (FC) layer. As is shown in Table~\ref{tb:visualenc}, when K=32, compared to V-TCN, using FC layer achieves 55\% less parameters while suffering a 1dB loss in SI-SDR and SDR. To achieve an extremely small model size, we use the FC layer for the visual encoder in this paper.

\subsection{Performance under causal configurations}
\label{causal}
The target application of LiMuSE is for resource-constrained devices. Therefore, it's necessary to discuss the online casual usage of the model. Here we explore the performance of LiMuSE under causal configurations. Table~\ref{tb:causal} indicates that the system still performs well under causal usage. The performance degradation could be the result of causal convolution and/or the cumulative layer normalization~\cite{luo2019conv}.

\subsection{Ablation study of LiMuSE}
\label{ablation}
We conducted an ablation study of LiMuSE to investigate the influence of lightweight designs and various modalities. 
\subsubsection{Effects of lightweight designs} \label{ssec:ablation1}
We investigate the effects of the Group Communication, Context Codec and the ultra-low bit quantization. The results are listed in Table~\ref{tb:ablation}.

The incorporation of Group Communication~(GC) and Context Codec~(CC) allows LiMuSE to maintain on par or even better performance under a suitable number of groups with a much smaller model size and less model complexity. We can see from the table that LiMuSE~(GC) slightly performs better than LiMuSE~(vanilla) when $K=16$ with only 0.37M parameters. LiMuSE~(GC+CC) with $K=16$ further improves the performance with 0.04M extra parameters but reduces the MACs from 11.77G to 7.52G compared to LiMuSE~(GC). It suggests that an appropriate number of groups may promote the inter-modal information exchange and achieve considerable improvements. When $K=32$, the number of parameters and the MACs of LiMuSE~(GC) and LiMuSE~(GC+CC) are further reduced, and the performance degradation of LiMuSE~(GC+CC) is very limited compared to LiMuSE~(vanilla). Intuitively, we can empirically conclude that the larger the value of K, the fewer the model parameters with some performance degradation in most cases. A trade-off can be considered between the model complexity and performance according to the needs of production.

Furthermore, we apply ultra-low bit quantization to compress the model size and achieves as much as 180$\times$ of compression ratio when $K=32$. In an extreme case, the model size of LiMuSE is compressed to 0.19MB. Combing the results of Table~\ref{tb:result}, the quantized LiMuSE still achieves competitive results over the audio-only baseline model TasNet~\cite{luo2019conv} and audio-visual baseline models AVMS~\cite{zhang2021avss} and AVDC~\cite{lu2019audio}. Though in \cite{tinywase2022}, the authors have shown that the ultra-low bit quantization technology can achieve on-par performance on audio-only TCN-based speaker extraction models, applying the ultra-low quantization to our audio-visual TCN-based LiMuSE model leads to some performance degradation when the weights are quantized to 3 bits. While deploying the model on resource-constrained devices, it is recommended that an appropriate quantization bit be chosen according to the trade-off between the performance and model size as is shown in Table~\ref{tb:compare_q}.

\subsubsection{Effects of the multimodal cues}
Based on the experimental results in Section~\ref{ssec:ablation1}, we investigate the effects of the visual cue and the voiceprint cue on the unquantized LiMuSE~(GC+CC). The results are listed in Table~\ref{tb:ablation_cue}. We can see from the table that the performance decreases after discarding the voiceprint cue or the visual cue, especially the visual cue. LiMuSE~(GC+CC-Vis) suffers from a larger performance gap compared to LiMuSE~(GC+CC-Vp). There is about a 10dB performance gap when $K=32$ and a 7dB performance gap when $K=16$ between LiMuSE and LiMuSE~(GC+CC-Vis). It suggests that the visual cue is more informative than the voiceprint cue. It can be further concluded that the voiceprint cue and the visual cue serve as complementary information for each other and introducing both of them into LiMuSE leads to the best performance compared with LiMuSE~(GC+CC-Vp) and LiMuSE~(GC+CC-Vis). 

\section{Conclusion}
\label{sec:conclusion}
In this paper, we propose a lightweight multi-modal speaker extraction model incorporating the voiceprint and visual features as complementary cues. We propose to use the GC-equipped TCN as the basic module to compress the model size, which is used in the Context Codec module and the separation network in LiMuSE. Furthermore, we explore the impact of the quantization technologies. In an extreme case, the model size of LiMuSE is compressed to 0.19MB. Our experiments show that LiMuSE achieves promising results on MC\_GRID dataset, and the quantized model achieves comparable performance with a compression rate as high as 180$\times$ compared with the baselines.

\section{ACKNOWLEDGMENTS}
\label{sec:ACKNOWLEDGMENTS}
This work was supported by the National Key Research and Development Program of China (2021ZD0201500).
\bibliographystyle{IEEEbib}
\bibliography{strings}

\end{document}